# STATISTICAL ANALYSIS OF NYC BUILDINGS AND WIND DAMAGES

**Elham Azimi, Bud Griffis**

**NYU Tandon School of Engineering**

## Abstract

The objective of this study is to determine the types of existing buildings that are at risk of falling debris based on height, age, construction classification, construction methods and materials and occupancy. This study focuses on elements that could become debris under high wind action and present a hazard to pedestrians, vehicles, and nearby structures. This study evaluated the particular building elements that might become Wind Generated Debris (WGD). This was accomplished by inspecting 500 buildings located in Manhattan that experienced wind-related incidents. The results illustrate that the building elements most likely to produce WGD are windows, followed by exterior fixtures, roof elements, stairs/sidewalk shed, and balcony elements, respectively. Consequently, FISP inspectors should pay particular attention to these elements, which have higher probabilities in causing incidents.

## Introduction and Background

Recent storms have spared New York City (NYC) from the maximum winds associated with tropical cyclones. As devastating as Hurricane Sandy was, not everything about the storm was unprecedented. Its 80 mile-per-hour (mph) peak wind gusts fell well short of other storms that have hit NYC, including Hurricane Carol in 1954 (up to 125-mph gusts) and Hurricane Belle in 1976 (up to 95-mph gusts). Previous storms also had much more precipitation. During Sandy, a scant inch of rain fell in some parts of New York, far less than the 5 inches of rain that fell during Hurricane Donna in 1960 or the 7.5 inches during the April 2007 nor'easter. With greater winds and more rain, Sandy could have had an even more serious impact on the areas that experienced the most devastation during the storm, including Staten Island, Southern Brooklyn, and South Queens. And while Sandy brought the full force of its impact during high tide at these southernmost areas of NYC, it hit the area around western Long Island Sound almost exactly at low tide.

Historically, NYC is no stranger to major storms. In 1821, a hurricane struck NYC, bringing winds of about 75 mph and a reported 13-foot storm surge that flooded Lower Manhattan as far north as Canal Street. In 1938, a storm known as the Long Island Express—which received its name because the fast-moving eye passed over Long Island—hit with no warning, leading to over 600 deaths, including 10 in NYC, while 100-mph wind gusts knocked out electricity north of 59th Street in Manhattan. In 1960, Hurricane Donna brought wind gusts of up to 90 mph and a 10-foot [above mean lower low water (**MLLW**)[1]] storm surge that caused extensive pier damage. In the last few decades, major storms have been forming in the North Atlantic with greater frequency. Storms are not the only climate threats New Yorkers face. The city is also vulnerable to other "extreme" events, such as heavy downpours, heat waves, droughts, and high winds. NYC is particularly vulnerable to high winds especially in connection with coastal storms. High winds down trees and collapse overhead utility lines, damaging property and causing power outages. At high enough speeds, winds can even damage buildings. Category 1 hurricanes have sustained wind speeds of at least 74 mph, and Category 2 hurricanes have sustained winds of 96 to 110 mph, far greater than Sandy's 80-mph wind speed at landfall in New Jersey. In fact, in 1954, Hurricane Carol brought sustained wind speeds of up to 100 mph to the New York area, causing extensive damage [4].

Hurricanes and tropical storms strike New York infrequently, relative to other types of coastal storms (generally arriving during hurricane season, which occurs from June 1 to October 31), and can produce large surges, heavy rains, and high winds. Nor'easters, by contrast, are cold weather storms that have strong northeasterly winds blowing in from the ocean ahead of them. Compared to hurricanes, nor'easters generally bring smaller surges and weaker winds. However, they can cause significant harm because they tend to last longer, resulting in extended periods of high winds and high water that can be sustained through one or more high tides.

---

[1] The United States' National Oceanic and Atmospheric Administration uses mean lower low water (**MLLW**), which is the average height of the lowest tide recorded at a tide station each day during the recording period (the National Tidal Datum Epoch - a 19-year period).

High winds are projected to pose a moderate risk to the building stock of NYC. While the NYC Panel on Climate Change (NPCC) does not provide specific projections for wind speeds, their projections do suggest an overall increase in the frequency of the most intense hurricanes, which are accompanied by high winds. Though the NYC Building Code already requires new buildings to implement standards protecting against top wind speeds associated with a Category 3 hurricane, older buildings that predate modern standards or have improperly installed and maintained external elements are vulnerable. Areas with open exposures—for instance, along the coasts—and older -one- and two-family homes are especially vulnerable. Additionally, all structures, including high-rise buildings, are susceptible to damage to façades, which can cause airborne debris during extreme wind events.

NYC's future wind risk profile in the face of climate change is uncertain. While current Building Code requirements are based on wind speed data from area airports (John F. Kennedy International Airport, LaGuardia Airport and Newark Liberty International Airport), a detailed mapping of the City's maximum wind profile could provide a much more accurate assessment of the risks that buildings face with potentially increased storm activity. Although current Building Code requirements are calibrated to withstand a Category 3 hurricane, as the climate changes this level will probably be seen as inadequate. To address this uncertainty and improve NYC's approach to protecting New Yorkers from wind risks, the City took the precautionary measure of amending the Building Code to clarify current wind-resistance specifications for façade elements, and it restricts the use of pea gravel and small dimension stone as ballast on roofs. The City, through the Mayor's Office of Long-Term Planning and Sustainability (OLTPS), implemented these Building Code changes in 2013. In addition, the City will expand the existing Department of Buildings (DOB) Façade Inspection Safety Program (FISP) for high-rise buildings to include rooftop structures and equipment [5]. Subject to available funding, the DOB will also initiate a study to more accurately map the wind profiles facing NYC's buildings across all five boroughs, identifying sites that face the greatest risk and recommending appropriate city responses. The goal was to commence this study in 2013, with completion expected in 2015, but contract action was delayed.

## Research Methodology

A set of activities was organized in order to determine the types of existing buildings that are at risk of falling debris based on height, age, construction classification, construction methods and materials and occupancy. These activities focused on elements that could become debris under high wind action and present a hazard to pedestrians, vehicles, and nearby structures.

A workflow for activities contributing to this study is shown in Figure 1.

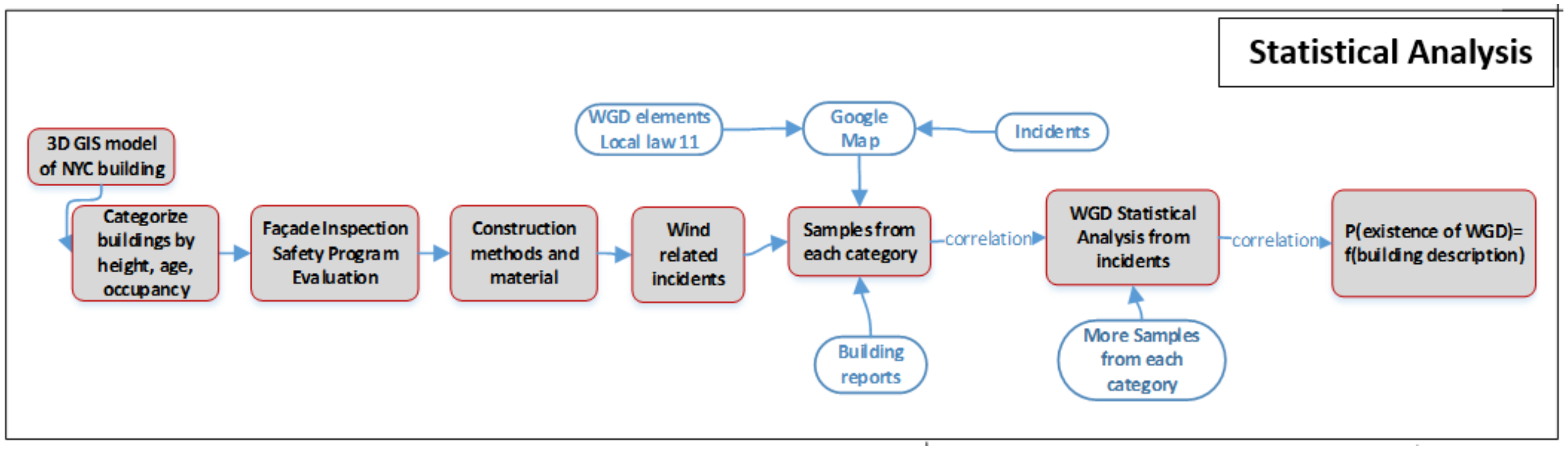

*Figure 1 Process to calculate the probability of wind generating debris*

### GIS Model of Building Types

Step 1 involved creating a 3-D Geographic Information System [6] model for NYC. This model covers nearly all the one million buildings that exist in the five boroughs of NYC. Buildings were colored by type of occupancy. A spatial analytical analysis was performed on this model further on in this study.

### Categorization of buildings

Step 2 involved categorizing the NYC buildings by height, age, occupancy and construction methods, and materials.

- Height

Buildings were first sorted into height categories. Height is classified into three main groups: low-rise, mid-rise and high-rise buildings. Low-rise buildings include two subcategories: single-story buildings and buildings up to six floors. Mid-rise buildings are those with 6 to 10 floors. High-rise buildings include three subcategories: buildings with 10 to 50 floors, buildings with 50 to 100 floors and buildings with more than 100 floors.

- Age and Governing Building Code

Buildings were then sorted by their age. A building's age was determined based on its year of completion. The age correlates with the building codes that govern the existing construction in NYC. Building codes were promulgated or revised in 1860, 1887, 1896, 1899, 1916, 1922, 1926, 1929, and then 1938, 1968, 2008 and 2014 [7].

The building codes from 1860 to 1916 are minimal and holistic in nature. These codes follow an integrated framework in which architectural, mechanical, structural and fire codes are combined together. Thus, changes in one item may impact the rest of the items. From 1938 to 2014, the evolution of construction management technology moved towards discrete systems whereby architectural, mechanical, structural and fire systems are described in different codes.

The main codes that govern existing buildings were key criteria in this study. Wind was not a design consideration in the building codes until the 1938 code. The codes were broken down into the following intervals: pre-1938, 1938 to 1968, 2008 to 2014, and post-2014. Even though architects and engineers considered wind forces while designing buildings prior to the 1938 code, they neglected them since the wind forces were not the controlling forces.

The following section describes the main aspects of each code.

**1938**: The focus of this version was to provide standards, provisions and requirements for safe and stable design, methods of construction and sufficiency of materials in structures constructed or demolished after January 1st, 1938. In addition, this code regulated the equipment, maintenance, use and occupancy of all structures and premises [8].

**1968**: This code specified minimum requirements and standards for the construction, alteration, repair, occupancy and use of new and existing buildings in the city of New York … All buildings to be maintained safely [9].

**2008**: Some jurisdictions like NYC developed their own building codes. This is the reason that this code is known as New York City Building Code (NYCBC).

However, due to ever increasing complexity and cost of developing and maintaining building regulations, virtually all municipalities in the country chose to adopt model codes recommended by International Code Council (ICC) instead. In 2008, NYC abandoned its proprietary 1968 Building Code in favor of a customized version of the International Building Code (IBC).

The IBC is founded on principles intended to establish provisions consistent with the scope of a building code that adequately protects public health, safety and welfare. These provisions include those that do not unnecessarily increase construction costs; provisions that do not restrict the use of new materials, products or methods of construction; and provisions that do not give preferential treatment to particular types or classes of materials, products or methods of construction [10].

**2014:** The purpose of this NYC construction code is to provide reasonable minimum load requirements and standards based on current scientific and engineering knowledge, experience and techniques, and the utilization of modern machinery, equipment, materials and forms and methods of construction. This code was updated in the interest of public safety, health, welfare and the environment, and with due regard for building construction and maintenance costs [11].

- Occupancy

According to various data sources, NYC's building stock is classified in different ways. The NYC Department of City Planning (DCP) [12] categorizes buildings using a “land use” or occupancy classification scheme. In this system, buildings are categorized by their building class and coded from 01 to 11 (Table 1). The buildings used in this study were classified according to the DCP’s system [13].

*Table 1 Land use types in NYC*

| CODES | DECODES |
|---|---|
| 01 | One & Two Family Buildings |
| 02 | Multi-Family Walk-Up Buildings |
| 03 | Multi-Family Elevator Buildings |
| 04 | Mixed Residential & Commercial Buildings |
| 05 | Commercial & Office Buildings |
| 06 | Industrial & Manufacturing |
| 07 | Transportation & Utility |
| 08 | Public Facilities & Institutions |
| 09 | Open Space & Outdoor Recreation |
| 10 | Parking Facilities |
| 11 | Vacant Land |

### Façade Inspection Safety Program (FISP) compliance of NYC buildings

The death of a 2-year-old girl hit by a piece of masonry that dislodged from an eighth-story windowsill stresses the importance of façade maintenance to public safety. In 1979, a similar incident left a Barnard College student dead and prompted the enactment of NYC Local Law 10 of 1980, the first in a series of groundbreaking façade safety ordinances that have since been the model for those in other major cities. In 1998, Local Law 11 tightened regulations to require inspection of all exterior walls, with evaluations performed via scaffolding for close inspection. Since then, the law has been updated and revised a number of times to improve its effectiveness. This NYC's Façade Inspection Safety Program (FISP), commonly known as Local Law 11/98, requires that owners of buildings taller than six stories must professionally inspect their buildings exterior walls and appurtenances once every five years [14].

### Past History of Wind-related Incidents

NYC 311 is the office that New Yorkers call to file a complaint [15] or service request regarding noise, transportation, public health and safety concerns, etc. This office is NYC's main source of government information and non-emergency services. Its mission is to provide the public with quick and easy access to all NYC government services and information while maintaining the highest possible level of customer service. It helps agencies improve service delivery by allowing them to focus on their core missions and manage their workload efficiently. It also provides insight into ways to improve City government through accurate, consistent measurement and analysis of service delivery citywide.

For this study, the 311 incidents reported to the 311 call [15] center that occurred in the five boroughs on windy days were evaluated statistically and analyzed. To ascertain these wind-related incidents incident data from 2010 to 2015 were considered. These data were then cross-correlated with the severely windy days as reported by the National Oceanic and Atmospheric Administration (NOAA) to extract only the windy-day related incidents. Table 2 shows the 311 complaints during windy days and their descriptions. These data were examined by borough to identify the types of complaints that were associated with potentially dangerous wind generated debris (WGD). It turned out that features of WGD complaints differ among the boroughs.

*Table 2 311 Complaints and Description*

| Complaint | Descriptor |
|---|---|
| Scaffold Safety | Suspended (Hanging) Scaffolds - No Pmt/Lic/Dangerous/Accident |
| BEST/Site Safety | Safety Netting/Guard Rails - Damaged/Inadequate/None (Over 6 Stories/75 Feet) |
| Cranes and Derricks | Crane/Suspension Scaffold - No Permit/License/Cert./Unsafe/Illegal |
| Special Projects Inspection Team (SPIT) | Sign - In Danger Of Falling |
| Stalled Sites | Stalled Construction Site |
| General Construction/Plumbing | Debris - Excessive |
| | Debris - Falling Or In Danger Of Falling |
| | Facade - Defective/Cracking (Ll11/98) |
| | Damage Assessment Request (Disaster) |
| | Site Conditions Endangering Workers |
| | Sidewalk Shed/Pipe Scafford - Inadequate Defective/None |
| | Safety Netting/Guard Rails - Damaged/Inadequate/None (6 Stories/75 Feet Or Less) |

The analysis provided critical insights as to what types of objects might fall during a windstorm. Local 311 investigation reports, DOB data and related reports and records of WGD, and records of previous damage considering the location, size, and cause of the damage were then compiled using the classification scheme in Table 2. Additional understanding of the characteristics of potential WGD came from various building condition reports based on hurricane incidents and high wind events. These reports have been obtained from several agencies and include:

- A Report on the City of New York's Response to Hurricane Sandy and the Path Forward---ONE CITY, REBUILDING TOGETHER [16],
- Hurricane Sandy Rebuilding Task Force by HUD [17],
- NYC Special Initiative for Rebuilding and Resiliency report: A Stronger, More Resilient New York [4] [5]
- New York Rising Year End Report [18].
- FEMA Hurricane Sandy reports [19].

Data for buildings of interest required for statistical analysis were collected from the latest version of Pluto2016 [12] and the 311 complaints [20]. Map PLUTO merges PLUTO tax lot data with tax lot features from the Department of Finance's Digital Tax Map (DTM), clipped to the shoreline. It contains extensive land use and geographic data at the tax lot level in ESRI ArcGIS shape format and database table format. Some of the Pluto features used by this study were the year of build, address, land use category, and the number of the floors. In order to have the geometry information of the buildings, these data were then combined with Building Footprints.

**Construction Materials and Methods**

Other features of buildings considered that related to WGD are listed in Table 3. The debris generating components include façade elements, roofing, window and balcony elements, façade material and protrusions such as ornamentation and other factors. These selected features are the same or similar to the critical examination that a New York State licensed architect or engineer conducts as set forth in RCNY 103-04. Using the façade of a building as an example, the potential WGD include exterior fixtures, flagpoles and signs. For roofing, the potential WGD include parapets, copings, roof facilities, railings, TV antennas, microwave towers and satellite dishes.

*Table 3 Summary of key factors considered in FISP study*

| category | Façade Fixture | Roofing | Balcony | Window | Stairs/Sidewalk shed | Façade Material |
|---|---|---|---|---|---|---|
| indicators | Camera | Parapets | Balcony enclosure | Window hardware | Stairs | Brick |
| | Light | Copings | Guard rails | Window light | Sidewalk shed | Stone |
| | Decorative elements | Facilities on roof | | Window railing | | Wood |
| | Buzzer | Railings | | Window air conditioners | | Concrete |
| | Antena | Chiemneys | | Flower boxes | | Curtain wall |
| | Flagpoles | TV antena | | | | Combination |
| | Signs | Microwave toweres | | | | other |
| | | Satelite dishes | | | | |

Data covering several buildings that experienced incidents on the windy days were identified. These buildings are widely distributed in the five NYC boroughs. Manhattan buildings were examined using Google Maps, and all potential falling objects were identified. After analyzing building components, features were ranked in their likelihood to have falling objects.

## Results and Discussion

The buildings of New York City were evaluated based on five main criteria. These criteria were Height, Age, Occupancy, Façade Inspection Safety Program, and Construction Materials and Methods. The results of these evaluations and statistical analyses are illustrated in this section.

### Height

Table 4 shows the number of NYC buildings categorized by their height. According to the Map Pluto 2016, there are 1,073,244 buildings in NYC. Queens has the largest number of buildings in the City with almost 43% of the total building stock, followed by Brooklyn with 30.6%, Staten Island with 13%, the Bronx with 9%, and Manhattan with only 4%.

Under the height category, buildings were broken up into separate groups. The first group includes two subcategories: single story buildings and buildings with 2 to 6 stories. There are 97,598 single story structures in NYC or 10% of the total number of buildings. Buildings with 2 to 6 stories make up almost 90% (89.7%) of the building stock. Both the mid-rise and high-rise buildings, which include buildings with greater than 6 stories, make up almost 1% of the building stock. As shown in Table 4, outside of Manhattan the majority of buildings have 1 to 2 floors. Manhattan buildings have mainly 4 to 5 floors; however, the majority of the buildings over 6 floors are located in Manhattan. Currently, there are only 98 buildings in NYC with 50 to 100 floors.  There are only two with over 100 floors, both located in the borough of Manhattan. The file used for categorizing building height was Map Pluto in which each floor height was considered to be 10 feet.

*Table 4 Categorizing NYC buildings based on height*

| Numbers of buildings in each category and borough | | | | | | sum | |
|---|---|---|---|---|---|---|---|
| floors | MN | BR | QU | SI | BX | | |
| 0<F<=1 | 1,243 | 23,085 | 40,855 | 20,660 | 11,755 | **97,598** | **1,060,366** |
| 1<F<=2 | 1,726 | **190,833** | **273,172** | **85,188** | **55,560** | **606,479** | |
| 2<F<=3 | 7,097 | 91,326 | 134,839 | 32,404 | 26,084 | **291,750** | |
| 3<F<=4 | 8,452 | 16,720 | 3,622 | 268 | 2,265 | **31,327** | |
| 4<F<=5 | **12,414** | 1,973 | 834 | 73 | 3,484 | **18,778** | |
| 5<F<=6 | 5,833 | 3,236 | 2,496 | 151 | 2,718 | **14,434** | |
| 6<F<=10 | 2,827 | 1,103 | 654 | 89 | 740 | **5,413** | **12,878** |
| 10<F<=50 | 5,212 | 833 | 457 | 65 | 798 | **7,365** | |
| 50<F<=100 | 94 | 3 | 1 | 0 | 0 | **98** | |
| F>100 | 2 | 0 | 0 | 0 | 0 | **2** | |
| Sum | 44,900 | 329,112 | 456,930 | 138,898 | 103,404 | **1,073,244** | |
| | | | | | | **1,073,244** | |

Figure 2 presents the NYC building distribution histogram based on height (number of the floors). Buildings with 1 to 2 floors make up more than half of all buildings in the City–approximately 56% of the building stock. Buildings with 2 to 3 floors make up 27% of all buildings. Single story buildings make up only 9% of all buildings. Finally, all other buildings make up less than 8% of the building stock cumulatively.

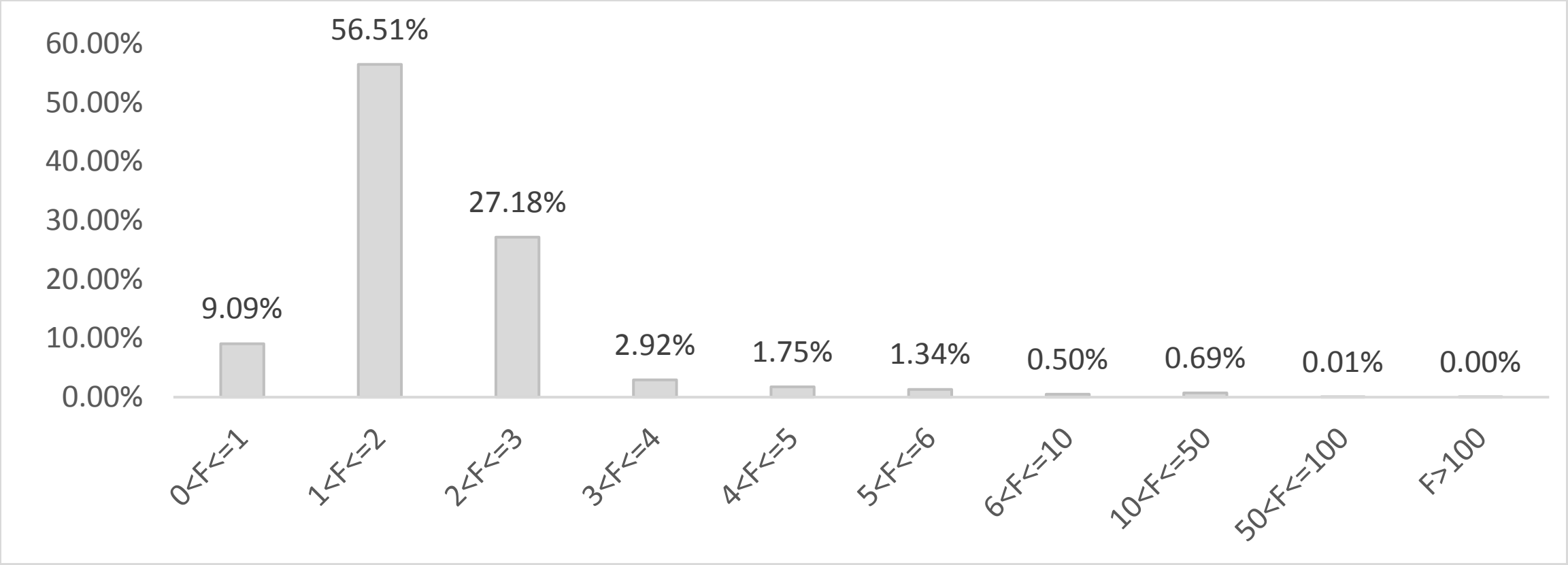

*Figure 2 NYC building distribution histogram based on number of the floors*

Figure 3 shows the City's building distribution histogram by number of floors and borough. In Queens, most buildings have 1 to 2 floors, followed by buildings with 2 to 3 floors and single-story structures, respectively. Queens also has the highest number of these building types. Brooklyn follows the same distribution pattern as Queens in terms of 1 to 2-floor buildings, 2 to 3-floor buildings and single-story structures, and has the second highest number of these building types. Staten Island and the Bronx follow the same distribution pattern as Queens and Brooklyn, and place third and fourth in terms of the number of these building types, respectively. Manhattan is the only borough in which the building distribution pattern is unique. In Manhattan, there are more buildings with 4 to 5 floors than any other type of building, followed by buildings with 3 to 4 floors and buildings with 2 to 3 floors, respectively. Also, Manhattan is the only borough that has buildings over 100 floors and 94 buildings with 50 to 100 floors.

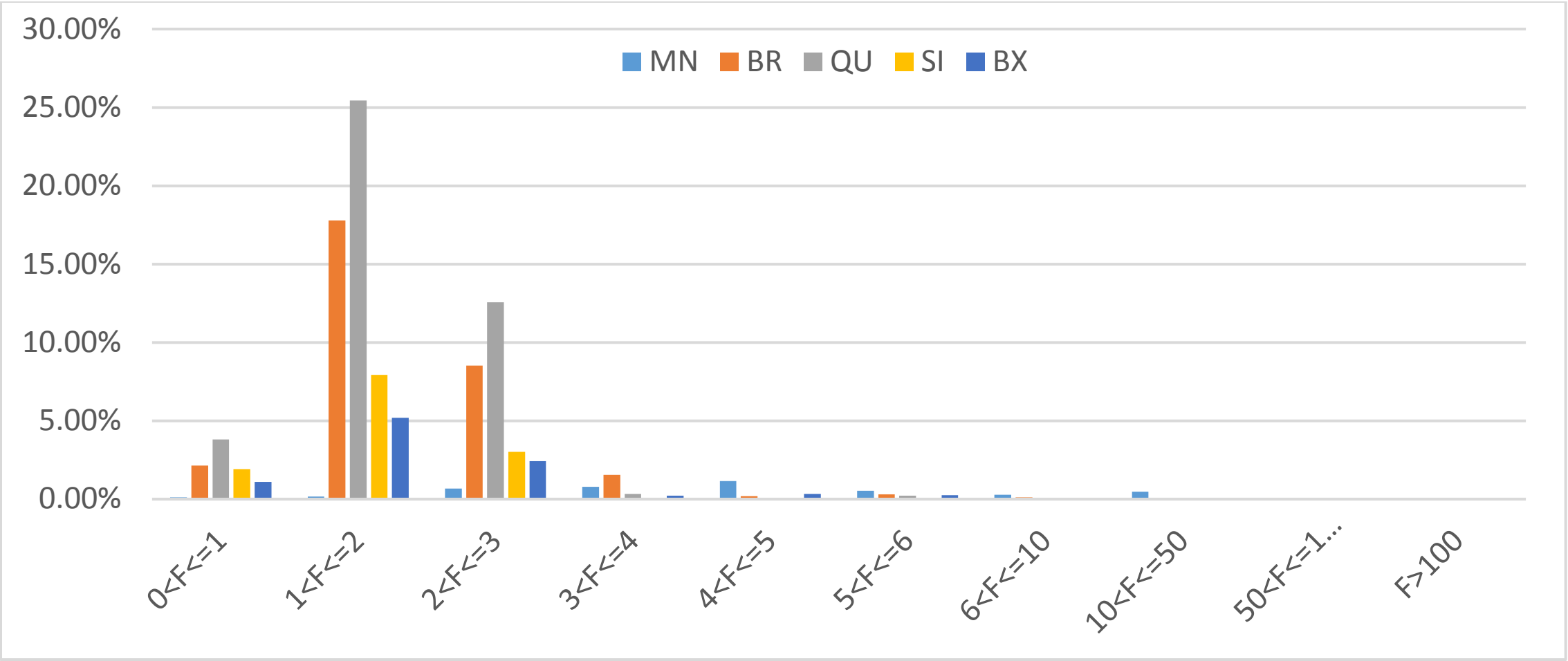

*Figure 3 NYC building distribution histogram based on number of the floors in 5 boroughs*

### Age

Table 5 and Figure 4 show NYC buildings classified by age. The number of the floors and the main driving building codes were compared. With the exception of Staten Island, the majority of the buildings in each borough were constructed before 1938. For example, in Manhattan more than 82% of buildings predate 1938. In Brooklyn and Queens, more than 76% and 54% of buildings were constructed prior to 1938, respectively. According to Table 5, there is a direct relationship between the height of buildings and their age in Brooklyn, Queens, and Bronx. In these boroughs, not only are the majority of the buildings classified as 1 to 2-floor buildings, but they were also built before 1938. Even though the majority of buildings in Manhattan were also built prior to 1938, 4 to 5-floor buildings were primarily built during that time.

*Table 5 NYC building categorization based on the age in 5 boroughs*

| | MN | | | | | | BR | | | | | | QU | | | | | |
|---|---|---|---|---|---|---|---|---|---|---|---|---|---|---|---|---|---|---|
| floors | <1938 | 1938<= <1968 | 1968<= <2008 | 2008<= <2014 | >=2014 | sum | <1938 | 1938<= <1968 | 1968<= <2008 | 2008<= <2014 | >=2014 | sum | <1938 | 1938<= <1968 | 1968<= <2008 | 2008<= <2014 | >=2014 | sum |
| 0<F<=1 | 584 | 412 | 199 | 14 | 3 | 1,212 | 9,488 | 10,690 | 2,164 | 211 | 37 | 22,590 | 9,544 | 27,837 | 2,635 | 164 | 12 | 40,192 |
| 1<F<=2 | 1,208 | 298 | 158 | 10 | 0 | 1,674 | **151,965** | **26,236** | **11,184** | 517 | 56 | 189,958 | **132,056** | **121,170** | **17,901** | **2,101** | **37** | 273,265 |
| 2<F<=3 | 6,491 | 252 | 276 | 6 | 0 | 7,025 | 72,987 | 6,590 | 10,399 | **1,132** | 98 | 91,206 | 105,976 | 16,235 | 11,823 | 1,081 | 30 | 135,145 |
| 3<F<=4 | 7,774 | 261 | 361 | 27 | 0 | 8,423 | 14,139 | 325 | 2,026 | 626 | **133** | 17,249 | 1,891 | 210 | 1,119 | 326 | 25 | 3,571 |
| 4<F<=5 | **11,769** | 412 | 194 | 23 | 0 | 12,398 | 1,184 | 109 | 591 | 243 | 32 | 2,159 | 415 | 90 | 228 | 119 | 7 | 859 |
| 5<F<=6 | 4,996 | 485 | 303 | 74 | 9 | 5,867 | 1,524 | 1,166 | 508 | 175 | 51 | 3,424 | 532 | 1,698 | 202 | 68 | 4 | 2,504 |
| 6<F<=10 | 1,968 | 309 | 491 | 140 | 22 | 2,930 | 256 | 281 | 467 | 175 | 97 | 1,276 | 55 | 287 | 198 | 108 | 14 | 662 |
| 10<F<=50 | 2,355 | **1,396** | **1,383** | **222** | **46** | 5,402 | 99 | 454 | 285 | 39 | 35 | 912 | 6 | 235 | 189 | 47 | 6 | 483 |
| 50<F<=100 | 11 | 5 | 60 | 12 | 6 | 94 | 0 | 0 | 0 | 1 | 2 | 3 | 0 | 0 | 1 | 0 | 0 | 1 |
| F>100 | 1 | 0 | 0 | 0 | 1 | 2 | 0 | 0 | 0 | 0 | 0 | 0 | 0 | 0 | 0 | 0 | 0 | 0 |
| SUM | 37,157 | 3,830 | 3,425 | 528 | 87 | 45,027 | 251,642 | 45,851 | 27,624 | 3,119 | 541 | 328,777 | 250,475 | 167,762 | 34,296 | 4,014 | 135 | 456,682 |

| SI | | | | | | BX | | | | | |
|---|---|---|---|---|---|---|---|---|---|---|---|
| <1938 | 1938<= <1968 | 1968<= <2008 | 2008<= <2014 | >=2014 | sum | <1938 | 1938<= <1968 | 1968<= <2008 | 2008<= <2014 | >=2014 | sum |
| 3,059 | 11,552 | 5,891 | 76 | 2 | 20,580 | 5,860 | 4454 | 1358 | 71 | 3 | 11,746 |
| **15,291** | **18,741** | **49,493** | **1,885** | **135** | 85,545 | **32,681** | **17,453** | 5,238 | 258 | 0 | 55630 |
| 18,796 | 1,329 | 12,000 | 200 | 7 | 32,332 | 13,301 | 4,525 | **8,018** | **458** | 4 | 26306 |
| 97 | 10 | 151 | 13 | 0 | 271 | 1,526 | 103 | 577 | 117 | **5** | 2328 |
| 51 | 13 | 2 | 2 | 0 | 68 | 3,330 | 40 | 89 | 28 | 0 | 3487 |
| 14 | 58 | 81 | 5 | 0 | 158 | 1,934 | 582 | 185 | 41 | 0 | 2742 |
| 15 | 57 | 16 | 1 | 0 | 89 | 110 | 344 | 233 | 58 | 1 | 746 |
| 0 | 4 | 60 | 2 | 0 | 66 | 30 | 429 | 325 | 75 | 0 | 859 |
| 0 | 0 | 1 | 0 | 0 | 1 | 0 | 0 | 0 | 0 | 0 | |
| 0 | 0 | 0 | 0 | 0 | 0 | 0 | 0 | 0 | 0 | 0 | |
| 37,323 | 31,764 | 67,695 | 2,184 | 144 | 139,110 | 58,772 | 27,930 | 16,023 | 1,106 | 13 | 103,844 |

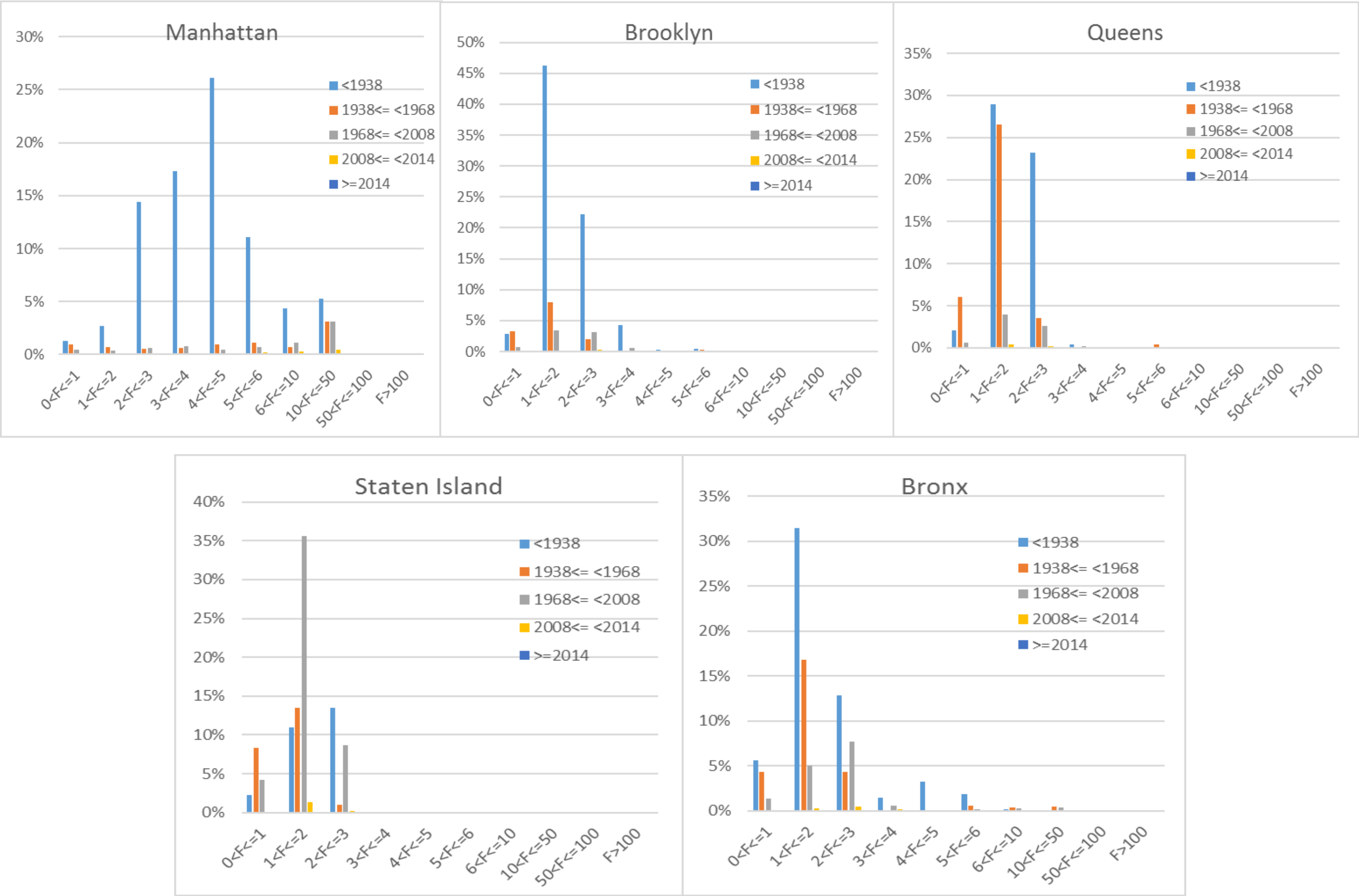

*Figure 4 NYC building distribution histogram based on age and number of the floors*

### Occupancy

*Table 6 Land use categories in NYC boroughs*

| land use code | Land Use description | Total | MN | BR | QN | SI | BX |
|---|---|---|---|---|---|---|---|
| 1 | One & Two Family Buildings | 753,560 | 4,048 | 201,068 | 364,421 | 120,958 | 63,065 |
| 2 | Multi-Family Walk-Up Buildings | 171,992 | 12,758 | 74,452 | 54,650 | 8,737 | 21,395 |
| 4 | Mixed Residential & Commercial Buildings | 59,519 | 12,606 | 25,925 | 14,306 | 1,923 | 4,759 |
| 5 | Commercial & Office Buildings | 24,318 | 4,813 | 6,542 | 7,202 | 2,493 | 3,268 |
| 8 | Public Facilities & Institutions | 16,550 | 2,954 | 5,454 | 3,947 | 1,319 | 2,876 |
| 3 | Multi-Family Elevator Buildings | 16,100 | 5,696 | 4,296 | 2,787 | 240 | 3,081 |
| 6 | Industrial & Manufacturing Buildings | 13,415 | 896 | 5,583 | 4,500 | 764 | 1,672 |
| 10 | Parking Facilities | 6,899 | 269 | 1,986 | 2,725 | 543 | 1,376 |
| 9 | Open Space & Outdoor Recreation | 2,556 | 213 | 430 | 681 | 603 | 629 |
| 7 | Transportation & Utility | 2,514 | 193 | 987 | 713 | 348 | 273 |
| 11 | Vacant Land | 1,053 | 116 | 396 | 297 | 160 | 84 |

As Table 6 and Figure 5 show, 70.5% of buildings in NYC are classified as one and two-family buildings, and 16.1% are categorized as multi-family walk-up buildings. These two types combined cover 86.6% of all buildings. The rest of the land use types cover only 14.4% of all buildings. However, the distribution of occupancy in different boroughs is not the same. In Brooklyn, Queens, Staten Island and the Bronx, one and two-family buildings constitute the most common land use types followed by multiple family walk-ups. Manhattan follows a different pattern; in Manhattan, the most common land use type is multi-family walk-ups, followed by mixed residential and commercial buildings and multi-family elevator buildings, respectively.

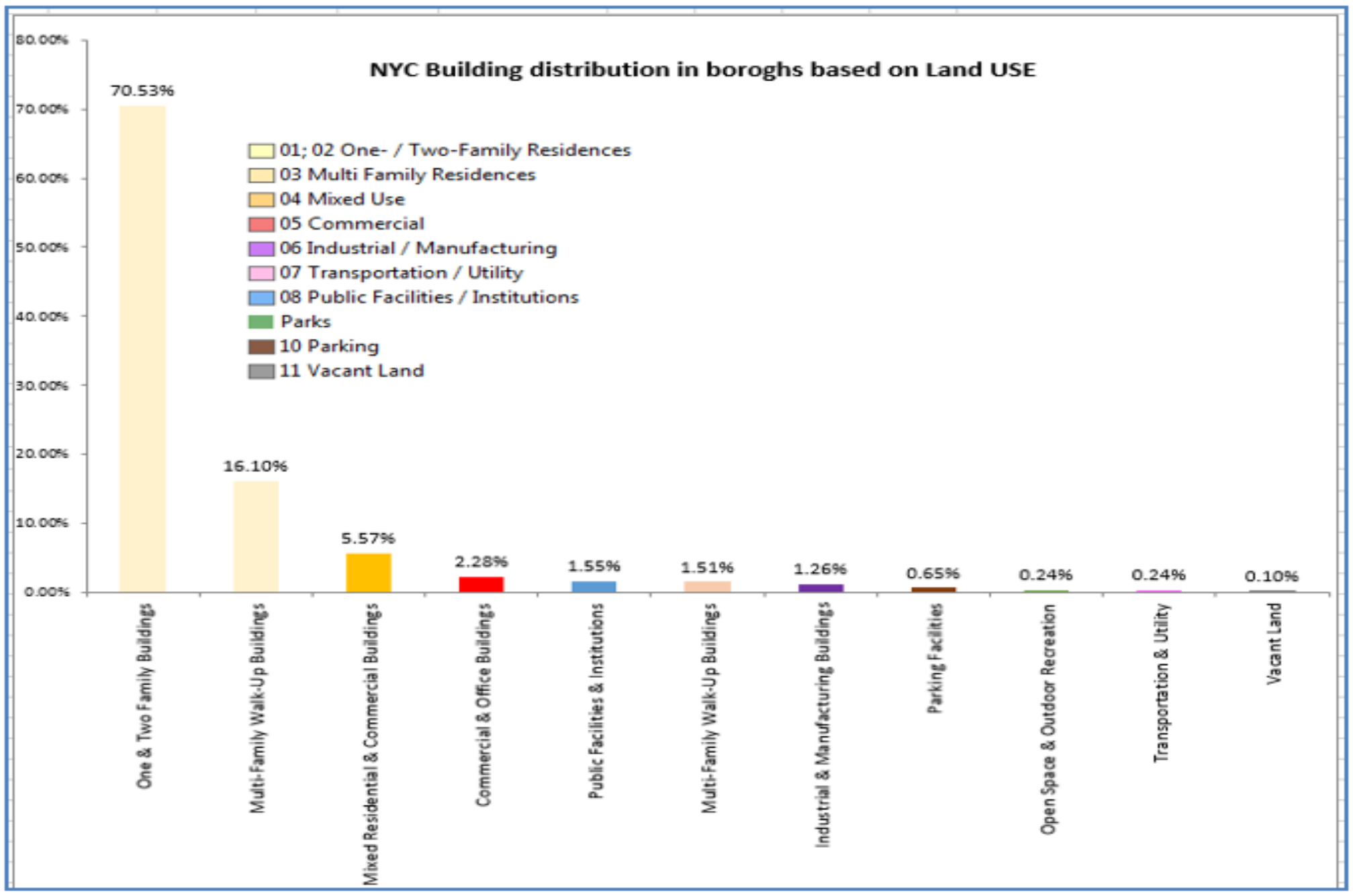

*Figure 5 NYC building distribution in boroughs based on Land Use*

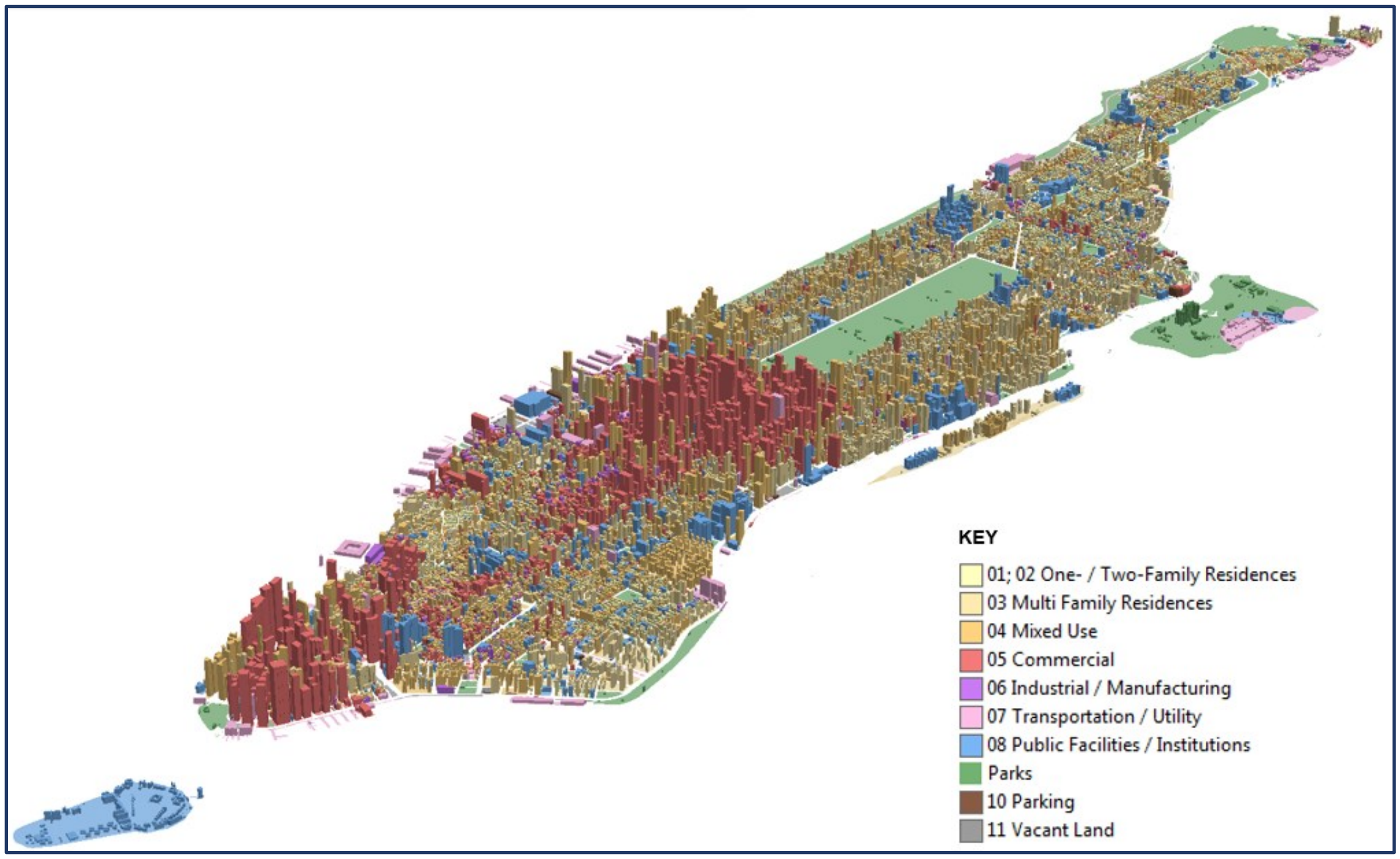

*Figure 6 NYC 3D mass model considering land use types*

Figure 6 shows a three-dimensional mass model of Manhattan that displays different land use types.

It should be mentioned that in some cases in Manhattan, buildings fitting under different land use categories could have a higher floor range than in the other boroughs. For example, in Manhattan, buildings with 1 to 6 floors are classified as one to two family residences. Additionally, while in all the other boroughs the maximum number of floors for a multi-family elevator building is 5 to 6 floors, the maximum number in Manhattan is 10 to 50 floors.

### FISP

The pie charts in Figure 7 indicate the building distribution in NYC by borough (on the left side) and the portion of NYC buildings that fall under FISP (on the right side). Manhattan buildings consist of 4% of total buildings in NYC. However, the buildings in this borough comprise 63% of FISP locations in NYC.

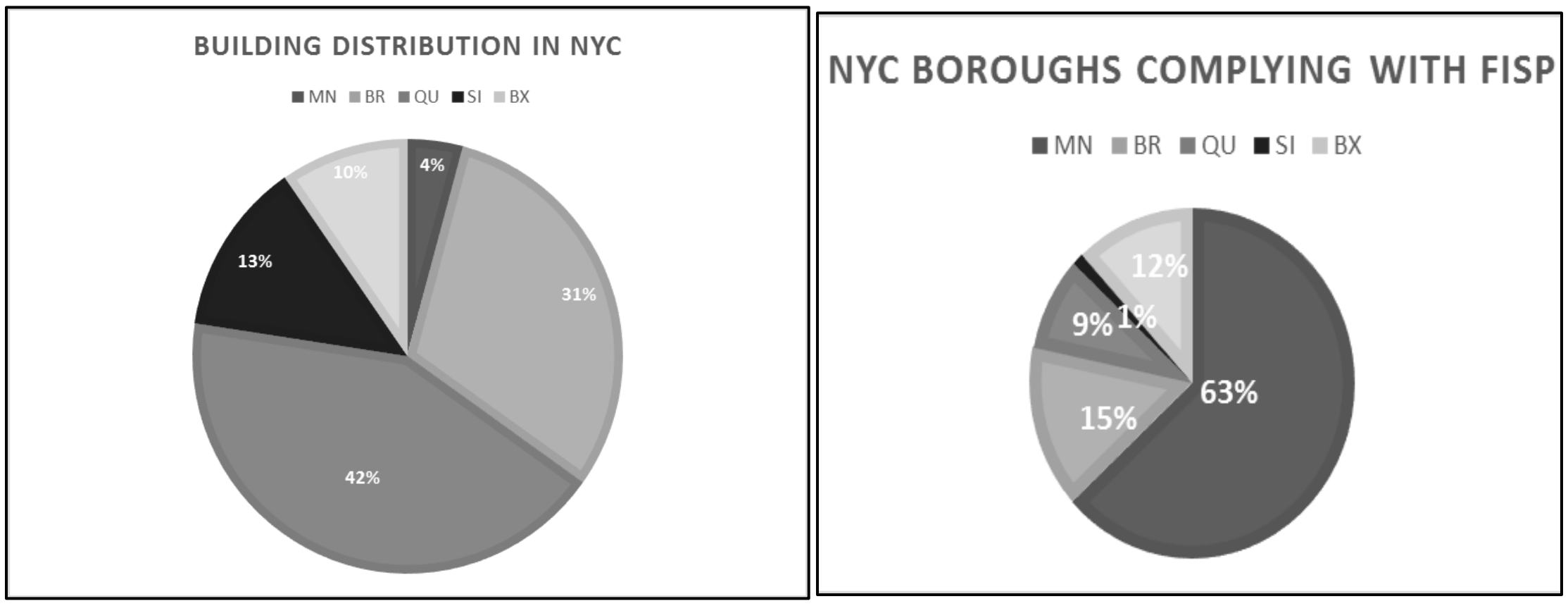

*Figure 7 Comparison between NYC buildings and the NYC buildings complying with FISP*

### Windy Day Incidents

Figure 8 shows the number of incidents that happened in NYC from 2010 to 2015 on the days that were reported windy by NOAA [21]. Out of 44,000 incidents, 2,300 were on windy days. After further analysis, 1,400 incidents related to existing buildings; the other incidents related to buildings that have been demolished or were under construction. Data were pulled from 311 [15] calls and reflect the complaints from residents of the existing buildings. The issue with 311 data is that they do not provide detailed information for each reported incident. For example, there is no data on the damage estimation (cost, material, injured or fatality rate) or the severity of each case.

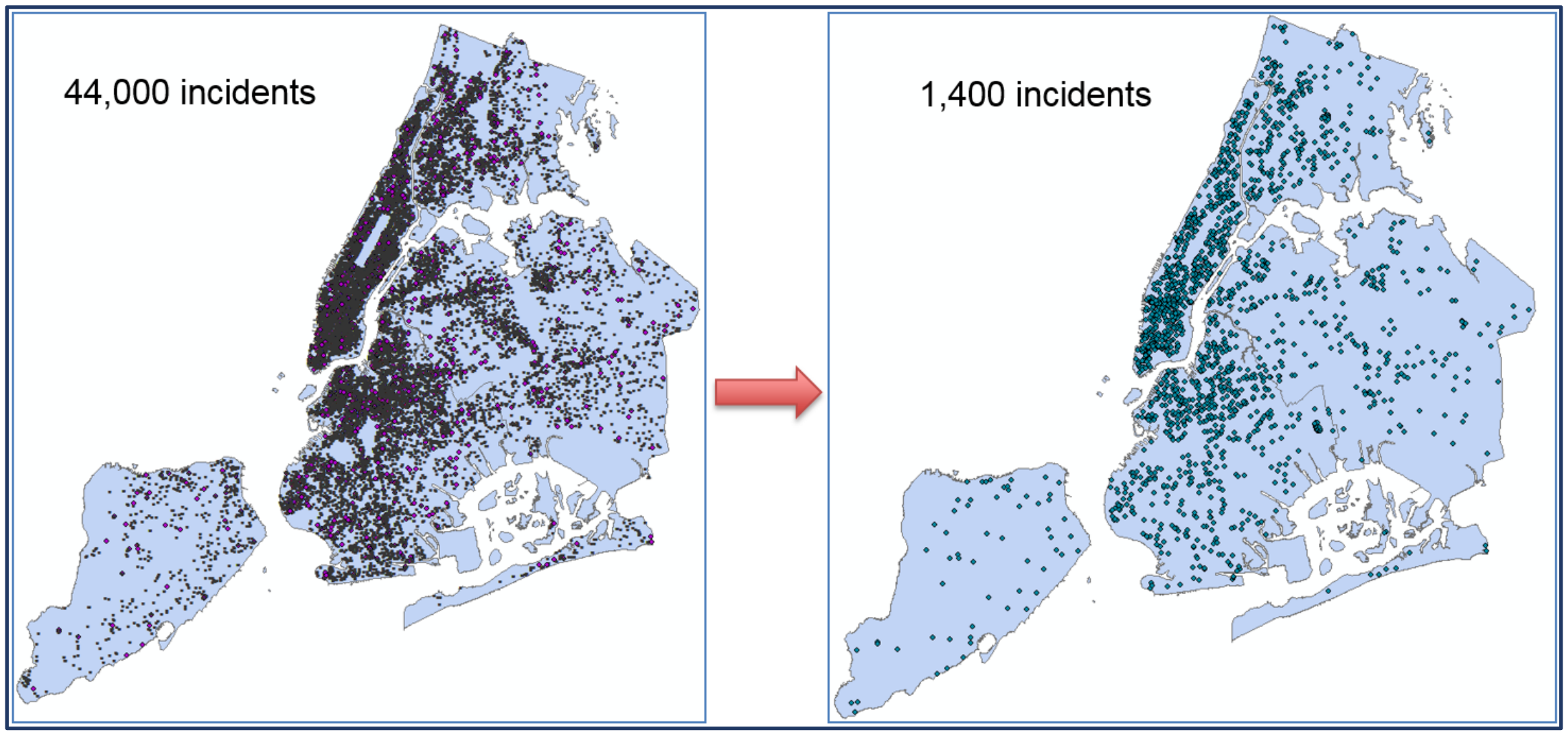

*Figure 8 Number of total wind incidents (left) and number of incidents relating to existing buildings (right)*

Figure 9 illustrates the results from categorizing different incidents from their description. The top three sources of complaints are debris falling or in danger of falling (30%), inadequate or defective sidewalk shed/pipe scaffold (24%), and suspended or hanging scaffold (16%).

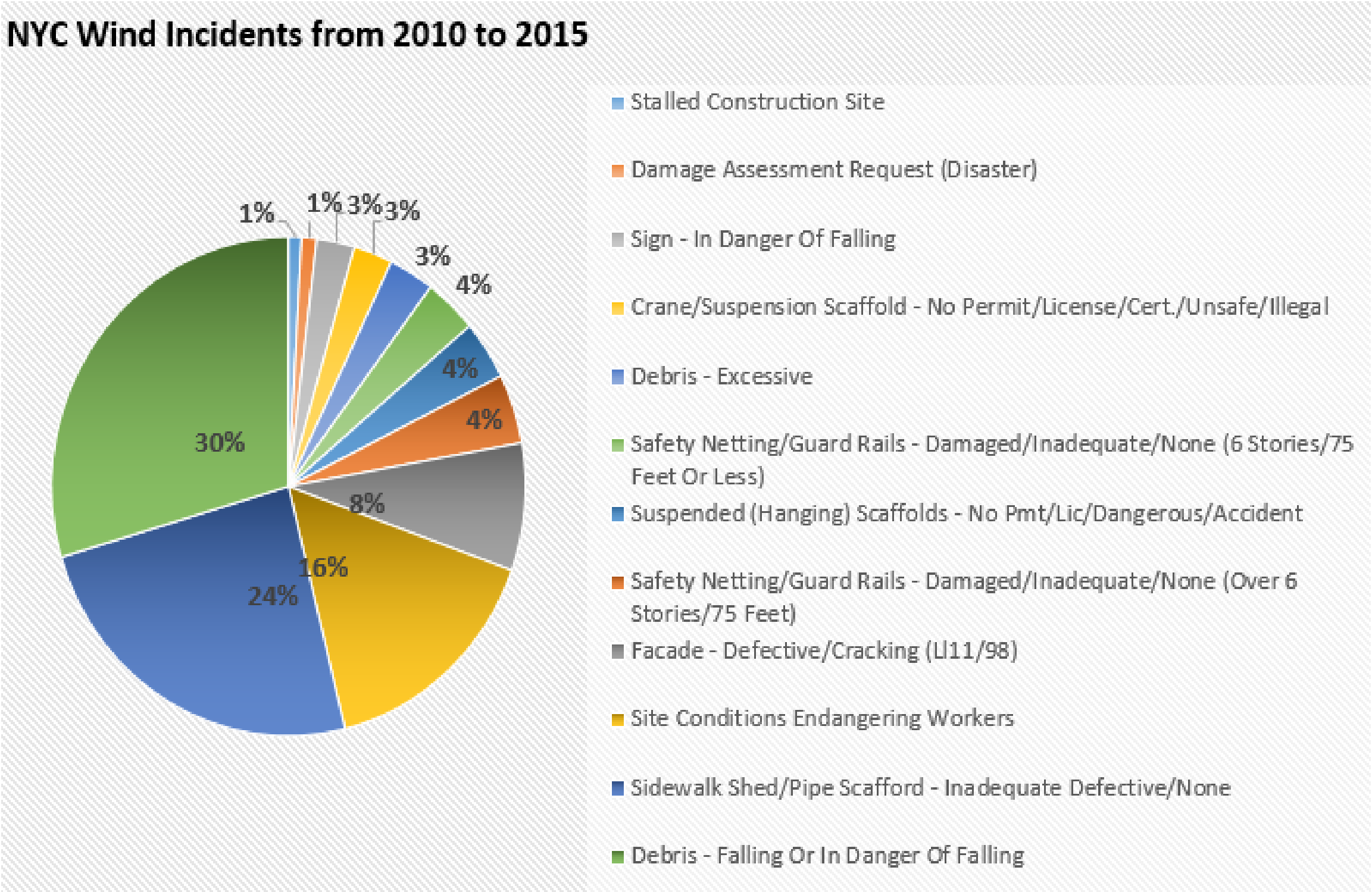

*Figure 9 NYC wind incidents from 2010-2015*

Although Figure 9 displays the cumulative incident categories deriving from 311 complaints [15] across NYC, this distribution differs within each borough. Figure 10 shows the number of different types of wind incidents that occurred during windy days in each borough from 2010 to 2015.

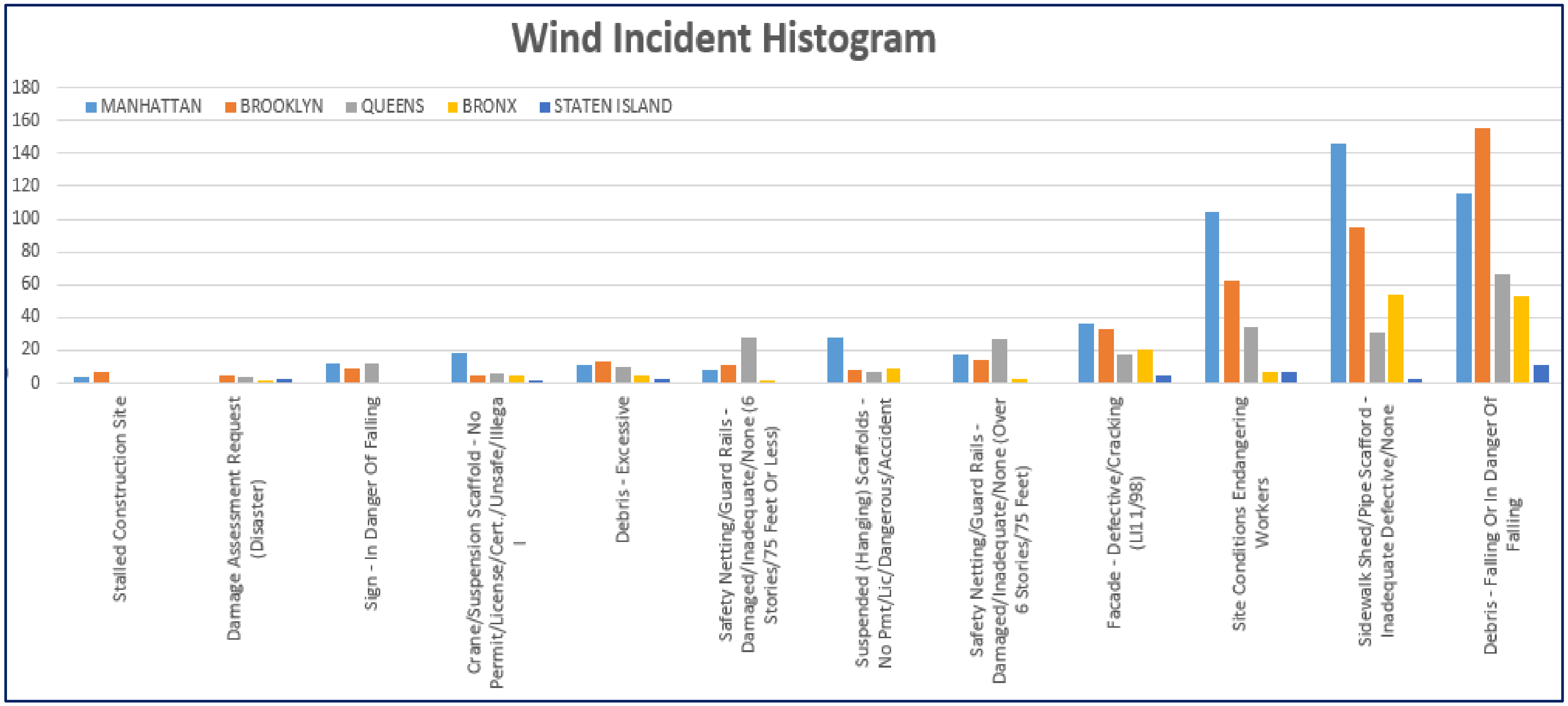

*Figure 10 NYC Wind Incidents Histogram in 5 boroughs*

### Construction materials and methods

Building appurtenances for the almost 1400 structures which experienced windy day incidents were used as case studies for this section. The vulnerable building features, which are listed in Table 3, were identified for each building. In all the boroughs except Manhattan, these identifications were made using Google Maps. Evaluation of the different parts of the façade was done for each individual building twice by different inspectors. The collected data were then compared, and the results reflected in this study. For Manhattan with around 500 incidents, data were gathered in person. The addresses were mapped and inspected by two groups of NYU students who visited each building during a two-week timespan. Due to the inaccessibility of the roof elements, these features were identified using Google Maps. In the result table (Table 8), if a building contains a component mentioned in Table 3, it was assigned the number 1. Likewise, if a building does not contain the component, it was assigned a 0. The average of each category (façade features, roofing, window and balcony data, stairs/sidewalk) was then calculated and used for further statistical analysis.

### Statistical analysis

The correlation between age and the probability of WGD was analyzed in this section. In addition, the likelihood of different building features causing WGD was evaluated and features were ranked based on the results.

- Relationship between the age of the building and the probability of having falling debris

The relationship between the year the buildings were built and their potential vulnerability to WGD is well described by a third degree polynomial function shown in Figure 11. This emphasizes the importance of exterior building maintenance, which is a critical principle highlighted in previous codes (1938, 1968, and 2008).

*Table 7 Probability of having falling debris vs age of the buildings of NYC*

| *Debris - Falling Or In Danger Of Falling* | *Year built* | *All the incidents within the duration* | *Debris falling within the duration* | *Age* | *Probability of having debris falling or in danger of falling* |
|---|---|---|---|---|---|
| Debris - Falling Or In Danger Of Falling | 1928 | 707 | 217 | 88 | 53.8% |
| Debris - Falling Or In Danger Of Falling | 1938 | 250 | 88 | 78 | 21.8% |
| Debris - Falling Or In Danger Of Falling | 1948 | 39 | 16 | 68 | 4.0% |
| Debris - Falling Or In Danger Of Falling | 1958 | 51 | 14 | 58 | 3.5% |
| Debris - Falling Or In Danger Of Falling | 1968 | 97 | 22 | 48 | 5.5% |
| Debris - Falling Or In Danger Of Falling | 1978 | 43 | 5 | 38 | 1.2% |
| Debris - Falling Or In Danger Of Falling | 1988 | 32 | 3 | 28 | 0.7% |
| Debris - Falling Or In Danger Of Falling | 1998 | 14 | 8 | 18 | 2.0% |
| Debris - Falling Or In Danger Of Falling | 2008 | 66 | 19 | 8 | 4.7% |
| Debris - Falling Or In Danger Of Falling | 2014 | 61 | 11 | 2 | 2.7% |
| TOTAL | | 1360 | 403 | | 100.0% |

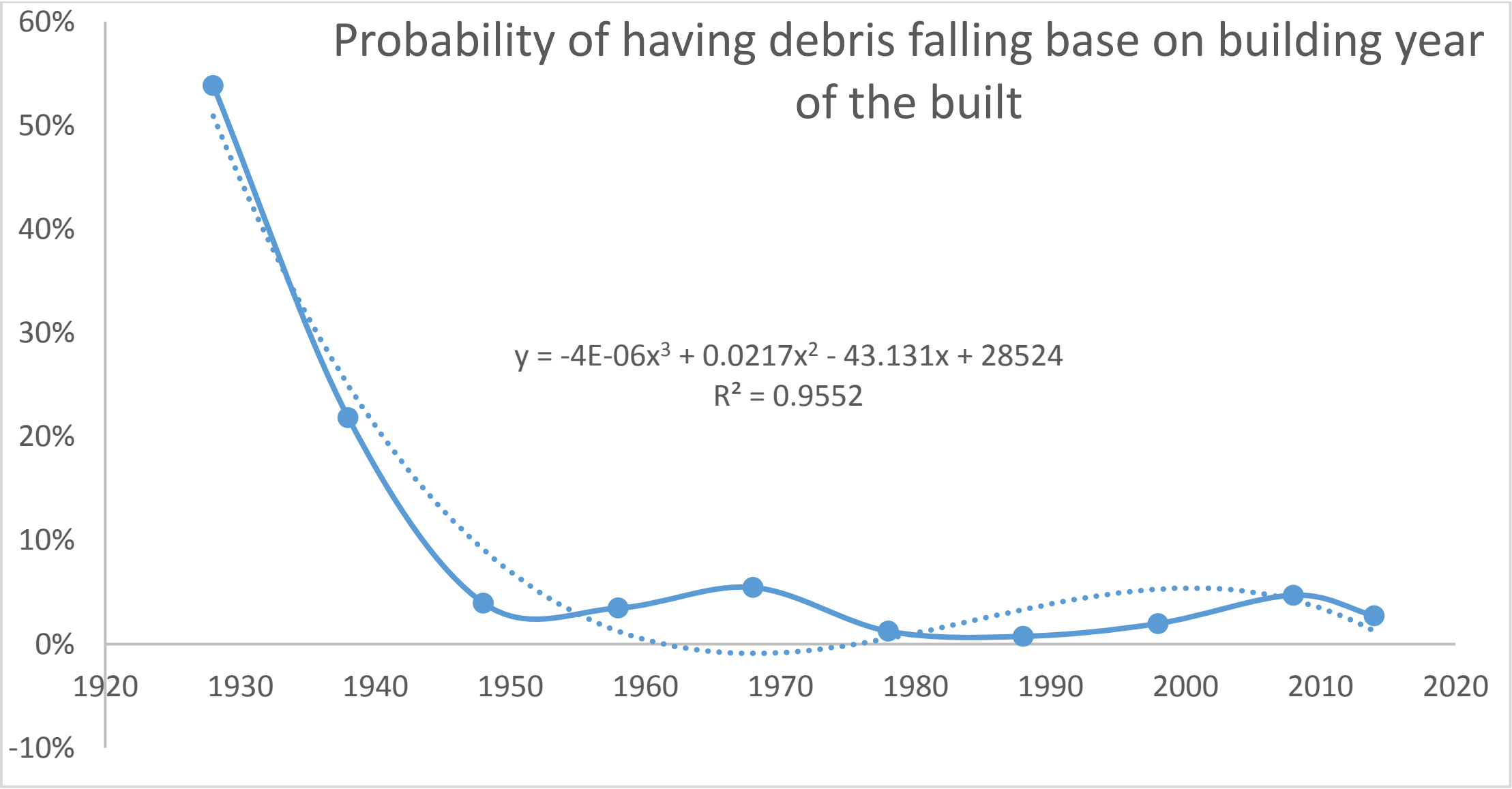

*Figure 11 Graph showing the probability of having falling debris and built year of the buildings*

- Ranking the elements involved in making the falling debris incident

The significance of the factors considered in FISP-building features was assessed. These elements and their subcategories were explained in Table 3. The results of the assessment were then confirmed by Analysis Of Variance (ANOVA) [22].

1-Input data:

Table 8 shows the results of the inspection of building features of 500 buildings in Manhattan. For each feature in the table, buildings with the feature receive a 1 and buildings without the feature receive a 0. Since the purpose of this section is to find the relationship between falling debris incidents and building features, only four types of the related incidents were evaluated. These types include: Debris-Excessive, Debris-Falling or in Danger of Falling, Façade-Defective/Cracking (LI11/98), and Sign in Danger of Falling. These incidents make up 175 of the total 500 incidents that occurred in Manhattan during windy days. Each row in Table 8 includes an address that was reported to 311 [15] and the elements of that building were inspected in person.

*Table 8 Snapshot of  inspection data from buildings experienced wind incidents in Manhattan*

| Match_addr | Descriptor | Façade Fixture | | | | | | | Roofing | | | | | | | | Balcony | | Window | | | | | Stairs/Sidewalk shed | |
|---|---|---|---|---|---|---|---|---|---|---|---|---|---|---|---|---|---|---|---|---|---|---|---|---|---|
| | | camera | light | decoration | buzzer | antena | flagpoles | signs | parapets | copings | facilities_roof | railings | Chimneys | TV_antena | Microwave_tower | satellite | balcony_enclosures | guard_rails | window_hardware | window light | window_railing | air_conditioners | flower_boxes | stairs | sidewalk |
| 12 E 72ND ST, 10021 | Debris - Falling Or In Danger Of Falling | 0 | 0 | 0 | 0 | 0 | 0 | 0 | 1 | 0 | 0 | 0 | 0 | 0 | 0 | 0 | 1 | 0 | 0 | 1 | 1 | 0 | 0 | 1 | 0 |
| 40 MARBLE HILL AVE, 1046 | Facade - Defective/Cracking (Ll11/98) | 1 | 0 | 0 | 0 | 0 | 0 | 0 | 0 | 0 | 0 | 0 | 0 | 0 | 0 | 1 | 0 | 0 | 0 | 0 | 1 | 1 | 0 | 1 | 0 |
| 101 COOPER ST, 10034 | Debris - Falling Or In Danger Of Falling | 1 | 0 | 0 | 0 | 0 | 0 | 0 | 1 | 0 | 0 | 1 | 0 | 0 | 0 | 0 | 0 | 0 | 1 | 0 | 1 | 1 | 0 | 1 | 0 |
| 600 W 207TH ST, 10034 | Debris - Falling Or In Danger Of Falling | 0 | 0 | 0 | 0 | 0 | 0 | 1 | 0 | 0 | 0 | 0 | 0 | 0 | 0 | 0 | 0 | 0 | 0 | 0 | 1 | 0 | 0 | 0 | 0 |
| 9 POST AVE, 10034 | Facade - Defective/Cracking (Ll11/98) | 1 | 0 | 0 | 0 | 0 | 0 | 0 | 1 | 1 | 0 | 0 | 0 | 0 | 0 | 0 | 0 | 0 | 0 | 0 | 1 | 1 | 0 | 1 | 0 |
| 121 DYCKMAN ST, 10040 | Debris - Falling Or In Danger Of Falling | 0 | 0 | 0 | 0 | 0 | 0 | 1 | 0 | 0 | 0 | 0 | 0 | 0 | 0 | 0 | 0 | 0 | 0 | 0 | 1 | 1 | 0 | 0 | 0 |
| 10 FAIRVIEW AVE, 10040 | Facade - Defective/Cracking (Ll11/98) | 0 | 0 | 0 | 0 | 0 | 0 | 0 | 1 | 0 | 0 | 0 | 0 | 0 | 0 | 0 | 0 | 0 | 0 | 0 | 1 | 1 | 0 | 1 | 0 |
| 607 W 180TH ST, 10033 | Debris - Falling Or In Danger Of Falling | 0 | 0 | 1 | 0 | 0 | 0 | 0 | 0 | 0 | 0 | 0 | 0 | 0 | 0 | 0 | 0 | 0 | 0 | 0 | 1 | 1 | 1 | 1 | 0 |
| 130 WADSWORTH AVE, 10( | Facade - Defective/Cracking (Ll11/98) | 0 | 0 | 1 | 0 | 0 | 0 | 0 | 0 | 0 | 0 | 0 | 0 | 0 | 0 | 0 | 0 | 0 | 0 | 0 | 1 | 1 | 0 | 1 | 0 |
| 700 W 172ND ST, 10032 | Debris - Falling Or In Danger Of Falling | 0 | 0 | 0 | 0 | 0 | 0 | 1 | 1 | 0 | 0 | 0 | 0 | 0 | 0 | 0 | 0 | 0 | 0 | 0 | 1 | 1 | 0 | 1 | 0 |
| 484 W 165TH ST, 10032 | Facade - Defective/Cracking (Ll11/98) | 1 | 0 | 0 | 1 | 0 | 0 | 0 | 0 | 0 | 0 | 0 | 0 | 0 | 0 | 0 | 0 | 0 | 1 | 0 | 1 | 1 | 0 | 1 | 0 |
| 421 W 162ND ST, 10032 | Debris - Excessive | 1 | 0 | 0 | 0 | 0 | 0 | 0 | 0 | 0 | 0 | 0 | 0 | 0 | 0 | 0 | 0 | 0 | 0 | 0 | 1 | 1 | 0 | 1 | 0 |
| 2785 8TH AVE, 10036 | Debris - Falling Or In Danger Of Falling | 1 | 0 | 1 | 0 | 0 | 0 | 1 | 1 | 0 | 0 | 0 | 0 | 0 | 0 | 0 | 0 | 0 | 1 | 0 | 1 | 0 | 0 | 0 | 0 |
| 635 RIVERSIDE DR, 10023 | Debris - Falling Or In Danger Of Falling | 1 | 0 | 0 | 0 | 0 | 0 | 0 | 0 | 0 | 0 | 0 | 0 | 0 | 0 | 0 | 0 | 0 | 0 | 0 | 1 | 1 | 1 | 0 | 0 |
| 400 W 145TH ST, 10031 | Debris - Excessive | 1 | 0 | 1 | 0 | 0 | 1 | 1 | 0 | 0 | 0 | 0 | 0 | 0 | 0 | 0 | 0 | 0 | 0 | 0 | 1 | 0 | 0 | 1 | 0 |
| 79 HAMILTON PL, 10031 | Debris - Falling Or In Danger Of Falling | 1 | 0 | 0 | 0 | 0 | 0 | 0 | 1 | 1 | 0 | 0 | 0 | 0 | 0 | 0 | 0 | 0 | 0 | 0 | 1 | 1 | 0 | 1 | 0 |
| 602 W 137TH ST, 10031 | Facade - Defective/Cracking (Ll11/98) | 1 | 0 | 1 | 0 | 0 | 0 | 0 | 0 | 1 | 0 | 0 | 0 | 0 | 0 | 0 | 0 | 0 | 0 | 0 | 1 | 1 | 0 | 0 | 0 |
| 630 W 136TH ST, 10031 | Debris - Falling Or In Danger Of Falling | 1 | 0 | 0 | 0 | 0 | 1 | 1 | 0 | 0 | 0 | 0 | 0 | 0 | 1 | 0 | 0 | 0 | 0 | 0 | 1 | 1 | 0 | 1 | 0 |
| 527 W 135TH ST, 10031 | Facade - Defective/Cracking (Ll11/98) | 1 | 0 | 0 | 0 | 0 | 0 | 0 | 0 | 0 | 0 | 0 | 0 | 0 | 0 | 1 | 0 | 0 | 0 | 0 | 1 | 1 | 0 | 1 | 0 |
| 135 W 143RD ST, 10030 | Facade - Defective/Cracking (Ll11/98) | 1 | 0 | 0 | 0 | 0 | 0 | 0 | 1 | 0 | 0 | 0 | 0 | 0 | 0 | 0 | 0 | 0 | 0 | 0 | 1 | 1 | 0 | 1 | 0 |
| 221 E 124TH ST, 10035 | Facade - Defective/Cracking (Ll11/98) | 1 | 0 | 0 | 0 | 0 | 0 | 0 | 0 | 0 | 0 | 0 | 0 | 0 | 0 | 0 | 0 | 0 | 0 | 0 | 1 | 1 | 0 | 1 | 0 |
| 328 E 86TH ST, 10028 | Sign - In Danger Of Falling | 0 | 0 | 0 | 0 | 0 | 0 | 0 | 1 | 0 | 1 | 0 | 0 | 0 | 0 | 0 | 1 | 1 | 0 | 0 | 0 | 1 | 0 | 0 | 0 |
| 216 E 84TH ST, 10028 | Debris - Excessive | 0 | 0 | 0 | 0 | 0 | 0 | 0 | 0 | 0 | 0 | 0 | 0 | 0 | 0 | 0 | 1 | 0 | 1 | 0 | 0 | 1 | 0 | 1 | 0 |

2-Methodology:

In order to prioritize the features in terms of their likelihood to create WGD, it was assumed that an incident caused by a subcategory of a feature impacted the likelihood of WGD from that feature as a whole. The probability of WGD depends on whether a building has one or more features. The more features a building has, the greater the probability of WGD. Since the feature categories and subcategories are mutually exclusive, the probabilities were added together. Table 9 shows the results of this assumption for some of the buildings.

*Table 9 Snapshot of data the portion of each category in making an incident*

| Match_addr | Descriptor | EX_ FIXTURE | ROOF | BAL | WINDOW | Stairs/Sidewalk shed |
|---|---|---|---|---|---|---|
| 12 E 72ND ST, 10021 | Debris - Falling Or In Danger Of Falling | 0.00 | 0.20 | 0.20 | 0.40 | 0.20 |
| 40 MARBLE HILL AVE, 10463 | Facade - Defective/Cracking (LI11/98) | 0.20 | 0.20 | 0.00 | 0.40 | 0.20 |
| 101 COOPER ST, 10034 | Debris - Falling Or In Danger Of Falling | 0.14 | 0.29 | 0.00 | 0.43 | 0.14 |
| 600 W 207TH ST, 10034 | Debris - Falling Or In Danger Of Falling | 0.50 | 0.00 | 0.00 | 0.50 | 0.00 |
| 9 POST AVE, 10034 | Facade - Defective/Cracking (LI11/98) | 0.17 | 0.33 | 0.00 | 0.33 | 0.17 |
| 121 DYCKMAN ST, 10040 | Debris - Falling Or In Danger Of Falling | 0.33 | 0.00 | 0.00 | 0.67 | 0.00 |
| 10 FAIRVIEW AVE, 10040 | Facade - Defective/Cracking (LI11/98) | 0.00 | 0.25 | 0.00 | 0.50 | 0.25 |
| 607 W 180TH ST, 10033 | Debris - Falling Or In Danger Of Falling | 0.20 | 0.00 | 0.00 | 0.60 | 0.20 |
| 130 WADSWORTH AVE, 100 | Facade - Defective/Cracking (LI11/98) | 0.25 | 0.00 | 0.00 | 0.50 | 0.25 |
| 700 W 172ND ST, 10032 | Debris - Falling Or In Danger Of Falling | 0.20 | 0.20 | 0.00 | 0.40 | 0.20 |
| 484 W 165TH ST, 10032 | Facade - Defective/Cracking (LI11/98) | 0.33 | 0.00 | 0.00 | 0.50 | 0.17 |
| 421 W 162ND ST, 10032 | Debris - Excessive | 0.25 | 0.00 | 0.00 | 0.50 | 0.25 |
| 2785 8TH AVE, 10036 | Debris - Falling Or In Danger Of Falling | 0.50 | 0.17 | 0.00 | 0.33 | 0.00 |
| 635 RIVERSIDE DR, 10023 | Debris - Falling Or In Danger Of Falling | 0.25 | 0.00 | 0.00 | 0.75 | 0.00 |
| 400 W 145TH ST, 10031 | Debris - Excessive | 0.67 | 0.00 | 0.00 | 0.17 | 0.17 |
| 79 HAMILTON PL, 10031 | Debris - Falling Or In Danger Of Falling | 0.17 | 0.33 | 0.00 | 0.33 | 0.17 |
| 602 W 137TH ST, 10031 | Facade - Defective/Cracking (LI11/98) | 0.40 | 0.20 | 0.00 | 0.40 | 0.00 |
| 630 W 136TH ST, 10031 | Debris - Falling Or In Danger Of Falling | 0.43 | 0.14 | 0.00 | 0.29 | 0.14 |
| 527 W 135TH ST, 10031 | Facade - Defective/Cracking (LI11/98) | 0.20 | 0.20 | 0.00 | 0.40 | 0.20 |
| 135 W 143RD ST, 10030 | Facade - Defective/Cracking (LI11/98) | 0.20 | 0.20 | 0.00 | 0.40 | 0.20 |
| 221 E 124TH ST, 10035 | Facade - Defective/Cracking (LI11/98) | 0.25 | 0.00 | 0.00 | 0.50 | 0.25 |
| 328 E 86TH ST, 10028 | Sign - In Danger Of Falling | 0.00 | 0.40 | 0.40 | 0.20 | 0.00 |

Sample addresses

The mean of each column was calculated and the results are shown in Figure 12. According to the results, window elements are the most likely to cause incidents, followed by exterior fixtures, roof elements, stairs/sidewalk shed, and balcony elements.

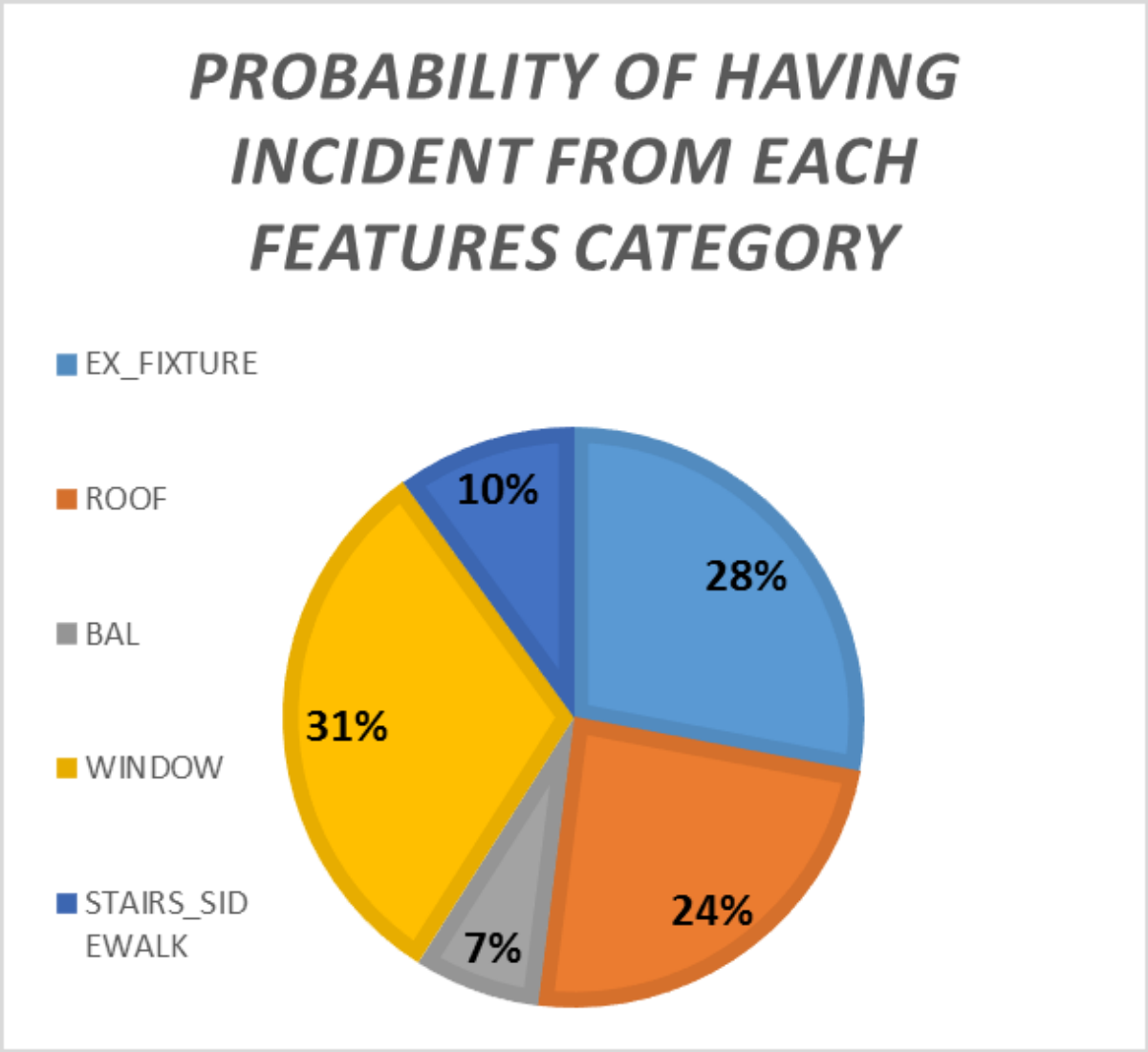

*Figure 12 Probability of having incident from each features category*

Table 10 illustrates the possible cases that were observed and the probabilities of having different features alone or with other features. In Table 10, cases that had a particular feature were assigned a value of 1, whereas cases without the feature were assigned a value of -1. So, there are two possibilities (1,-1) for each building feature. There are five building features here ($2^5$) which make thirty two possible cases. The features were compared alone and together in table 10. The probability column illustrates the probability of the +1-features. When more than one feature are +1, this shows the probability of having these +1-features together.

This table was used for Design of Experiment (DOE) in ANOVA [22] analysis as well. The name "analysis of variance" is based on the approach in which the procedure uses variances to determine whether the means are different. Multi-factor Analysis of Variance was used for this case in Minitab [23]. The procedure works by comparing the variance between group means versus the variance within groups, which produces the F-Value.

3-Output:

Table 10 are illustrating the ANOVA results.

*Table 10 Results of having different features in making the incident*

| cases | Façade Fixture | Roofing | Balcony | Window | Stairs/Sidewalk shed | Probability |
|---|---|---|---|---|---|---|
| 1 | -1 | -1 | -1 | -1 | -1 | 0.00 |
| 2 | -1 | -1 | 1 | -1 | -1 | 0.07 |
| 3 | -1 | -1 | -1 | -1 | 1 | 0.09 |
| 4 | -1 | -1 | 1 | -1 | 1 | 0.16 |
| 5 | -1 | 1 | -1 | -1 | -1 | 0.26 |
| 6 | 1 | -1 | -1 | -1 | -1 | 0.29 |
| 7 | -1 | -1 | -1 | 1 | -1 | 0.29 |
| 8 | -1 | 1 | 1 | -1 | -1 | 0.34 |
| 9 | -1 | 1 | -1 | -1 | 1 | 0.35 |
| 10 | 1 | -1 | 1 | -1 | -1 | 0.36 |
| 11 | -1 | -1 | 1 | 1 | -1 | 0.36 |
| 12 | 1 | -1 | -1 | -1 | 1 | 0.37 |
| 13 | -1 | -1 | -1 | 1 | 1 | 0.38 |
| 14 | -1 | 1 | 1 | -1 | 1 | 0.42 |
| 15 | 1 | -1 | 1 | -1 | 1 | 0.45 |
| 16 | -1 | -1 | 1 | 1 | 1 | 0.45 |
| 17 | 1 | 1 | -1 | -1 | -1 | 0.55 |
| 18 | -1 | 1 | -1 | 1 | -1 | 0.55 |
| 19 | 1 | -1 | -1 | 1 | -1 | 0.58 |
| 20 | 1 | 1 | 1 | -1 | -1 | 0.62 |
| 21 | -1 | 1 | 1 | 1 | -1 | 0.63 |
| 22 | 1 | 1 | -1 | -1 | 1 | 0.64 |
| 23 | -1 | 1 | -1 | 1 | 1 | 0.64 |
| 24 | 1 | -1 | 1 | 1 | -1 | 0.65 |
| 25 | 1 | -1 | -1 | 1 | 1 | 0.66 |
| 26 | 1 | 1 | 1 | -1 | 1 | 0.71 |
| 27 | -1 | 1 | 1 | 1 | 1 | 0.71 |
| 28 | 1 | -1 | 1 | 1 | 1 | 0.74 |
| 29 | 1 | 1 | -1 | 1 | -1 | 0.84 |
| 30 | 1 | 1 | 1 | 1 | -1 | 0.91 |
| 31 | 1 | 1 | -1 | 1 | 1 | 0.93 |
| 32 | 1 | 1 | 1 | 1 | 1 | 1.00 |

*Table 11 ANOVA analysis*

Factor Information

| Factor | Type | Levels | Values |
|---|---|---|---|
| A | Fixed | 2 | -1, 1 |
| B | Fixed | 2 | -1, 1 |
| C | Fixed | 2 | -1, 1 |
| D | Fixed | 2 | -1, 1 |
| E | Fixed | 2 | -1, 1 |

Analysis of Variance

| Source | DF | Adj SS | Adj MS | F-Value | P-Value |
|---|---|---|---|---|---|
| A | 1 | 6410.5 | 6410.52 | 8.33367E+08 | 0.000 |
| B | 1 | 4763.9 | 4763.86 | 6.19301E+08 | 0.000 |
| C | 1 | 410.4 | 410.41 | 53353462.50 | 0.000 |
| D | 1 | 7495.8 | 7495.78 | 9.74451E+08 | 0.000 |
| E | 1 | 724.7 | 724.66 | 94206118.50 | 0.000 |
| Error | 26 | 0.0 | 0.00 | | |
| Total | 31 | 19805.2 | | | |

| | |
|---|---|
| 9.74E+8 | D |
| 8.33E+8 | A |
| 6.19E+8 | B |
| 0.97E+8 | E |
| 0.53E+8 | C |

In Table 11, sources A to E refer to the building features. A refers to Façade Fixture, B to Roofing, C to Balcony, D to Window, and E to Stairs/Sidewalk Shed. In order to understand the results, the P-Values and F-Values from Table 11 were examined. Sorting the F-Value from highest to lowest illustrates the importance of the elements.

According to these results, window elements are most likely to cause incidents, followed by exterior fixtures, roof elements, stairs/sidewalk shed, and balcony elements. These results confirm the results of Table 10.

Figure 13 shows the controlling graphs to confirm the results. These graphs include:

- Patterns in normal probability plot

The normal probability plot of the residuals should approximately follow a straight line.

- Patterns in residuals versus fitted values plot

The residuals in the plot should be evenly spread across fitted values.

- Patterns in residuals versus order plot

The residuals in the plot should fluctuate in a random pattern around the center line.

According to the definitions, the graphs shown in Figure 13 follow the pattern that they should.

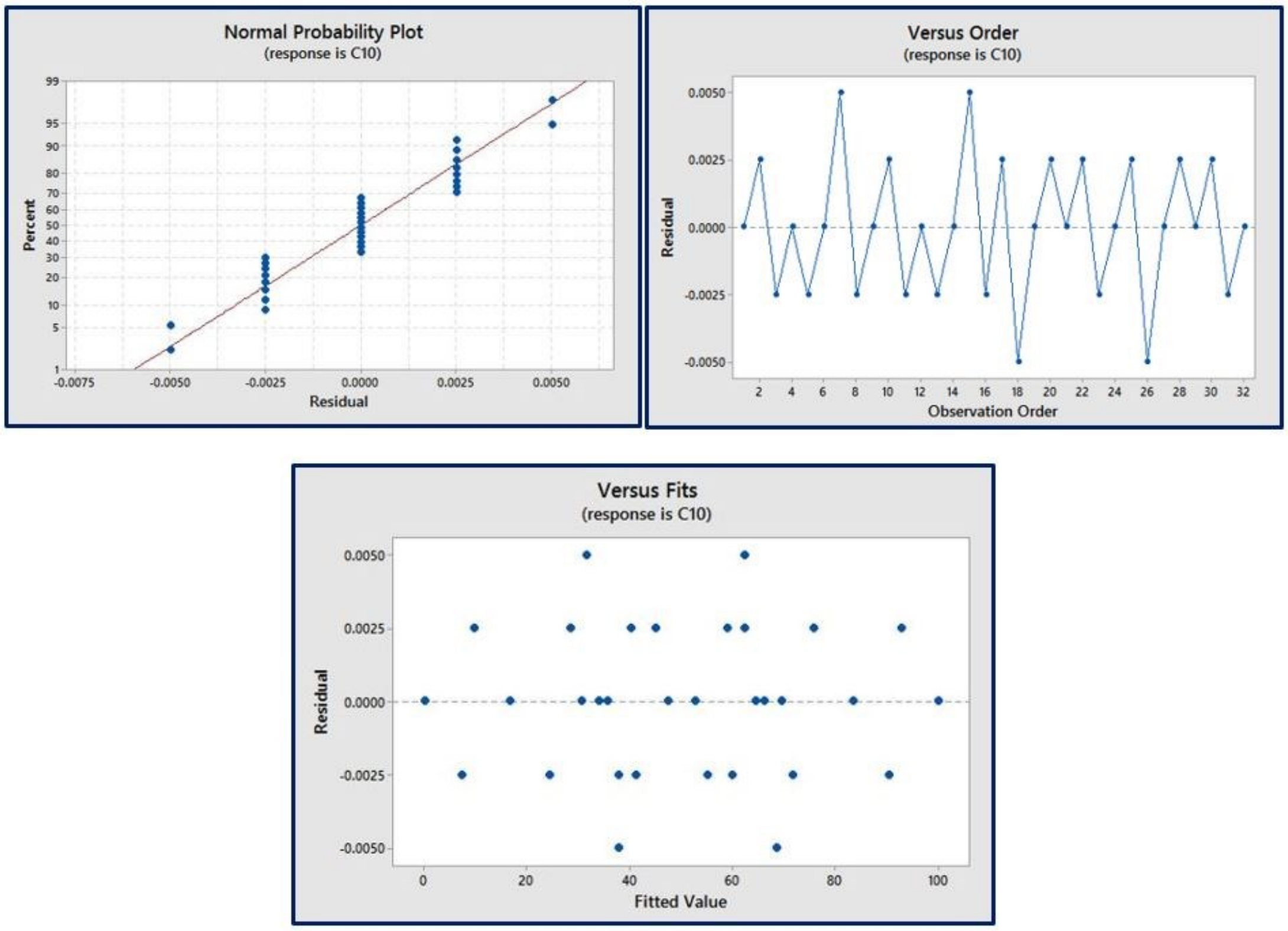

*Figure 13 Validity of the results*

## Conclusion

NYC is particularly vulnerable to high winds especially in connection with coastal storms. NPCC projections suggest an overall increase in the frequency of the most intense hurricanes, which are accompanied by high winds. At high enough speeds, winds can damage buildings. Though the NYC Building Code already requires new buildings to implement standards protecting against top wind speeds associated with a Category 3 hurricane, older buildings that predate modern standards or have improperly installed and maintained external elements are vulnerable.

Falling debris due to windy conditions is a particular concern as it threatens the lives and safety of New Yorkers. To determine the likelihood of WGD from existing buildings in NYC, this study started by categorizing the buildings using factors such as height, age and governing building code, FISP compliance, construction methods and materials and occupancy. Then, the history of wind-related incidents was analyzed in the five boroughs from 2010 to 2015 on days that were reported windy by NOAA. Of the total 44,000 wind-related incidents, roughly 1400 were related to existing buildings. The other incidents related to under construction or demolished buildings.

The relationship between the year each building was constructed and its potential vulnerability to WGD was assessed and turned out to be well described by a third degree polynomial function. This emphasizes the importance of exterior building maintenance, which a critical principle is highlighted in previous codes (1938, 1968, and 2008).

Next, this study evaluated the particular building elements that might become

WGD. This was accomplished by inspecting 500 buildings located in Manhattan that experienced wind-related incidents. The results illustrate that the building elements most likely to produce WGD are windows, followed by exterior fixtures, roof elements, stairs/sidewalk shed, and balcony elements, respectively. Consequently, FISP inspectors should pay particular attention to these elements, which have higher probabilities in causing incidents.

The incident data used in this study were pulled from 311 calls [15] and reflect the complaints from residents of existing buildings. It is strongly suggested that NYC 311, the DOB, and insurance companies who have the related data track the costs (direct and indirect) of each and every incident in an open database, providing the opportunity for academia and engineers to navigate the reasons behind similar incidents in buildings and provide adequate provisions for upcoming events.

# HIGHLIGHTS

Statistical Analysis

- In this study a new set of topics for categorization of the buildings was introduced: these categories include height, age, and occupancy. Buildings were also categorized considering the FISP criteria. Although the necessity of having this way of classification was needed for the largest city of United States, no records were found to describe this topic in the literature review. Therefore, the results of this study can be of a great use of urban designers, engineers, and governmental stakeholder's in macro-decision making for the City.
- The assessment of the relationship between the year each building was constructed and its potential vulnerability to WGD was take place in this study for the first time. The findings emphasize the importance of exterior building maintenance, which is a critical principle highlighted in previous codes (1938, 1968, and 2008).
- This study evaluated the particular building elements that might become WGD by inspecting the buildings located in Manhattan that experienced wind incidents in the past. Consequently, FISP inspectors should pay particular attention to these elements, which have higher probabilities in causing incidents.